\documentstyle[12pt]{article}
\def\cN{{\cal{N}}}

\begin{document}

\begin{titlepage}
\noindent
hep-th/9903149
\hfill ITP--UH--07/99 \\

\vskip 3.0cm

\begin{center}

{\Large\bf TOWARDS A STRINGY EXTENSION OF}\\

\bigskip

{\Large\bf SELF--DUAL SUPER YANG--MILLS}

\vskip 2.5cm

{\Large 
Anton Galajinsky\footnote{
Permanent address:\\ \phantom{XX}
Department of Mathematical Physics, 
Tomsk Polytechnical University, 634034 Tomsk, Russia.}
\ \ and \  
Olaf Lechtenfeld}

\vskip 1.0cm

{\it Institut f\"ur Theoretische Physik, Universit\"at Hannover}\\
{\it Appelstra\ss{}e 2, 30167 Hannover, Germany}\\
{E-mail: galajin, lechtenf@itp.uni-hannover.de}

\end{center}
\vskip 3.0cm

\begin{abstract}
Motivated by the search for a space--time supersymmetric
extension of the $N{=}2$ string, we construct a particle model 
which, upon quantization, describes (abelian) self--dual super 
Yang--Mills in $2{+}2$ dimensions. The local symmetries of 
the theory are shown to involve both world--line supersymmetry 
and kappa symmetry. 

\end{abstract}

\vfill


\end{titlepage}

\noindent
{\bf 1. Introduction}\\
Recent progress~\cite{dl,jlp} in a stringy description of $(2{+}2)$--dimensional
self--dual Yang--Mills and self--dual gravity triggered by investigations of 
$N{=}2$ strings (see~\cite{reviews} for reviews containing further references) 
stimulates interest in the construction of supersymmetric extensions. 
The first formulation of such a supersymmetric self--dual 
system~\cite{siegel2} has been obtained from a constraint system known earlier
as the $d{=}10$ ``$ABCD$ superstring'' (or superparticle)~\cite{siegel1}, by 
truncating it to a self--dual superspace.

Since the proposed set of constraints fails to form a closed algebra
\cite{pope1}, it was suggested to examine a smaller but closed subset.
Two corresponding $N{=}2$ string models were 
constructed~\cite{pope1,pope2}. However, their
massless states, although consisting of a scalar and
a spinor, do not interact according to self--dual super Yang--Mills (SDSYM)
or self--dual supergravity (SDSG)\footnote{A successful stringy 
extension of $\cN{=}1$ SDSG was achieved by de Boer
and Skenderis~\cite{deboer} who combined a conventional 
$N{=}2$ string in the right--moving sector with a 
Green--Schwarz--Berkovits type sigma model
\cite{gsb} in the left--moving sector.}.

It is well known that quantum field theories appear not only as 
low--energy limits of string theories but also upon quantization
of point--particle models.
Guided by the connection of the $\cN{=}1$ superparticle to $\cN{=}1$ $d{=}10$ 
super Yang--Mills theory, it is tempting to investigate similar relations
in a superspace based on $R^{2,2}$ which we call Kleinian superspace.
The knowledge of the zero mode structure for that case is likely 
to clarify the construction of a full superstring.

Yet, the conventional $N{=}2$ string~\cite{ov1} does not possess
space--time supersymmetry (see, however, Ref.~\cite{bkl}).
Its point analogue~\cite{howe} is the $N{=}2$ case of a spinning
particle with $N$--extended local world--line supersymmetry~\cite{sparticle}.
Upon quantization in $d{=}4$ such a model describes two massless 
irreps of the Poincar\'e group, labelled by helicities $\pm\frac N2$. 
For $N{=}2$ these are just the self--dual resp. anti--self--dual
parts of ordinary Yang--Mills theory.

Even though the $N{=}1$ spinning string turns into the superstring
after the GSO treatment, space--time supersymmetry is not present
in spinning particle or $N{=}2$ string models and, thus, requires
an enlargement of these theories.
The incorporation of rigid supersymmetry in the $d{=}4$ spinning particle
(which provides also a $\kappa$ symmetry)
lifts the original Poincar\'e irreps to super Poincar\'e irreps of
superhelicity $\pm\frac N2$~\cite{vanholten1,vanholten2}. 
One immediately concludes that the only possibility involving the 
$\cN{=}1$ SDSYM multiplet is to choose $N{=}1$ 
(see also Ref.~\cite{vanholten1}).

Our consideration suggests that a stringy extension of 
SDSYM may have to be doubly sypersymmetric, i.e.
possessing both local world--sheet supersymmetry and $\kappa$ 
symmetry (for an earlier consideration of doubly supersymmetric models 
see e.g. Ref.~\cite{double}). 
It is the purpose of the present letter to construct 
a doubly supersymmetric particle model in Kleinian superspace which 
describes abelian SDSYM when covariantly quantized.

\bigskip

\noindent
{\bf 2. Abelian SDSYM}\\
Let us begin by reminding the reader of the spinor notation in Kleinian 
flat space--time in order to fix our notation.
The covering group of the Lorentz group in $R^{2,2}$ factorizes as
\begin{equation}
Spin(2,2)\ =\ SL(2,R)\times SL(2,R)' \quad,
\end{equation}
and we distinguish Weyl spinors with respect to the two factors by
employing undotted and dotted early Greek indices, respectively.
The standard convention is applied when contracting spinor indices,
i.e. $\psi^2=\psi^\alpha\psi_\alpha$ and 
${\tilde\phi}^2={\tilde\phi}_{\dot\alpha}{\tilde\phi}^{\dot\alpha}$.
Contrary to the case of $SL(2,C)$, the fundamental $SL(2,R)^{(\prime)}$ 
spinor representations are Majorana--Weyl and, hence, not related by 
complex conjugation.  We take all spinors to be real whereas
sigma matrices are chosen to be imaginary. 

Irrespective of the space--time signature, 
abelian SYM is described by a single Majorana spinor 
superfield $(W_\alpha,\tilde {W}_{\dot\alpha})$.
A convenient way to incorporate the self--duality condition
in Kleinian superspace is to drop the anti--chiral spinor superfield strength
and work with $W_\alpha$ alone~\cite{ketov}.
Thus, our starting point is
\begin{equation}\label{Weqs}
\tilde {D}_{\dot\alpha} W_\beta (x,\theta,\tilde\theta)\ =\ 0
\qquad{\rm and}\qquad
D^\alpha W_\alpha (x,\theta,\tilde\theta)\ =\ 0 \quad,
\end{equation}
where $D_\alpha=\partial_\alpha + 
{\tilde\theta}^{\dot\alpha}\partial_{\alpha\dot\alpha}$ and
${\tilde D}_{\dot\alpha}=-\partial_{\dot\alpha}-
\theta^\alpha\partial_{\alpha\dot\alpha}$ are the usual covariant 
derivatives. For $R^{2,2}$ this is consistent because 
$W_\alpha$ is real and automatically fulfills the Majorana--Weyl 
condition.\footnote{The integrability conditions for (\ref{Weqs}) 
imply that $W_\alpha$ obeys the Weyl equation in the external 
index.}

The first of equations~(\ref{Weqs}) is 
just the chirality condition which specifies the field content 
of the problem and reduces the component expansion of 
$W_\alpha$ to
\begin{eqnarray}
W_\alpha (x,\theta,\tilde\theta) &=& \lambda_\alpha(x)+
\theta^\beta F_{\beta\alpha}(x)+ \theta^2 \phi_\alpha (x) \nonumber\\ &&\ +\
\theta^\beta\tilde\theta^{\dot\beta}\partial_{\beta\dot\beta}\lambda_\alpha(x)
+{\textstyle{\frac 12}} 
\theta^2\tilde\theta^{\dot\beta}\partial_{\beta\dot\beta}{F^\beta}_\alpha(x)
+{\textstyle{\frac 14}} \theta^2{\tilde\theta}^2 \Box\lambda_\alpha(x) \quad.
\end{eqnarray}
The second condition in (\ref{Weqs}) is the equation of motion. Decomposing 
$F_{\alpha\beta}=F_{(\alpha\beta)}+\epsilon_{\alpha\beta} F$ one finds  
\begin{equation}\label{eom1}
F(x)\ =\ 0 \quad, \qquad \phi_\alpha(x)\ =\ 0 \quad,
\end{equation}
\begin{equation}\label{eom2} 
\partial^{\alpha\dot\alpha} F_{(\alpha\beta)}(x)\ =\ 0 \quad, \qquad
\partial^{\alpha\dot\alpha} \lambda_\alpha(x)\ =\ 0 \quad.
\end{equation}

The first of Eqs.~(\ref{eom1}) implies that $F_{\alpha\beta}$ are the
components of a self--dual field strength; the first of Eqs.~(\ref{eom2})
is nothing but the corresponding Bianchi identity. 
The latter guarantees (locally) the existence of a self--dual vector potential
(we symmetrize with weights),
\begin{equation}
F_{(\alpha\beta)}\ =\ 
{\partial_{(\alpha}}^{\dot\gamma} A_{\beta)\dot\gamma}
\qquad{\rm with}\qquad
0\ =\ \partial_{\gamma(\dot\alpha}\ {A^\gamma}_{\dot\beta)} \quad.
\end{equation}

All equations above are invariant under abelian gauge transformations
(acting only on $A$) as well as under (1,1) supersymmetry:
\begin{equation}
\delta\lambda_\alpha\ =\ \epsilon^\beta F_{(\alpha\beta)} 
\qquad{\rm and}\qquad
\delta A_{\alpha\dot\alpha}\ =\ 2\,\lambda_\alpha\ \tilde\epsilon_{\dot\alpha} 
\quad\longrightarrow\quad \delta F_{(\alpha\beta)}\ =\ 2\,
{\partial_{(\alpha}}^{\dot\alpha}\lambda_{\beta)}\ \tilde\epsilon_{\dot\alpha}
\end{equation}
where $\epsilon^\alpha$ and ${\tilde\epsilon}^{\dot\alpha}$ are
two independent (real) Majorana--Weyl spinors of opposite chirality.
It is worth mentioning that the supersymmetry algebra closes 
(modulo gauge transformations) only on--shell,
i.e. {\it with the use of the self--duality condition}.

We shall go on to construct a spinning (super) 
particle model which, when covariantly quantized, yields
Eqs.~(\ref{Weqs}) (in their complexified form). It is known that 
the quantum wave function of a conventional spinning particle is realized 
in terms of four component Dirac (and not Weyl) spinors~\cite{berezin}. 
It is therefore convenient to rewrite our starting point~(\ref{Weqs}) as
\begin{equation}
(1+\gamma_5)_{AB}\ W_B\ =\ 0 \quad,\qquad
{\gamma^n}_{AB}\ \partial_n W_B\ =\ 0 \quad,
\end{equation}
\begin{equation}\label{majorana}
\phantom{XXX}
\tilde{D}_{\dot\alpha} W_A\ =\ 0 \quad,\qquad
\partial^{\alpha\dot\alpha} D_\alpha W_A\ =\ 0 \quad,
\end{equation}
with $W_A=\left({W_\alpha \above0pt \tilde W^{\dot\alpha}}\right)$ 
a Majorana spinor and 
$\gamma_n=\left({0\ \sigma_n \above0pt {\tilde\sigma}_n\ 0}\right)$
the Dirac matrices in Kleinian flat space--time.\footnote{
Note that ${\gamma_n}^T=C^{-1}\gamma_n C$ with $C=\gamma_0\gamma_1$
in Kleinian space--time.} 
The first two equations define 
an on--shell chiral fermion, while the remaining ones govern
supersymmetry.

\bigskip 

\noindent
{\bf 3. Chiral fermion model}\\
A technique for obtaining the massless Dirac equation from 
first--quantized particle mechanics has been known for a long time. 
It suffices to start with the action~\cite{berezin}
\begin{equation}\label{oldaction}
S\ =\ \int d\tau \Bigl\{ \frac 1{2e}(\dot x^n+i\chi\psi^n)^2+
i\psi_n {\dot\psi}^n \Bigr\} \quad,
\end{equation}
where $\psi^n$ and $\chi$ are anticommuting real vector and scalar variables, 
respectively.
Quantization modifies the Grassmann algebra to a Clifford algebra, so that 
the $\psi^n$ will be represented by gamma matrices.
Incorporation of the chirality condition $1+\gamma_5=0$
into the model~(\ref{oldaction}) turns out to be more involved. 
As was shown in Ref.~\cite{cortes1}, naively implementing the classically 
analogous constraint~\cite{howe} by adding a Lagrange multiplier term
\begin{equation}
\delta S\ =\ \int d\tau\ \phi\ (1+\epsilon^{abcd}\psi_a\psi_b\psi_c\psi_d)
\end{equation}
is classically inconsistent\footnote{
The approach followed in Ref.~\cite{gitman} suffers from the same problem
\cite{cortes1}.}
since taking the body of the constraint equation implies $0=1$. 
Interestingly, the contradiction seems to be avoided when one imposes only the
{\it time derivative\/} of the classical contraint. 
On the quantum level, however, one has to be more inventive.
An elegant solution is provided through
an extension of the system by a conventional fermionic oscillator.\footnote{
A similar extension for the massive Dirac particle has been employed
in Ref.~\cite{cortes2}.}
So, let us consider the following action,
\begin{equation}\label{newaction}
S\ = \int^{\tau_2}_{\tau_1}\!\! d\tau\ \Bigl\{ {\textstyle{\frac 1{2e}}}
(\dot x^n+\omega\epsilon^{nmkl}\psi_m\psi_k\psi_l+\chi\psi^n)^2
+\psi_n {\dot\psi}^n+\bar\xi\dot\xi +\phi\,
(i\bar\xi\xi-{\textstyle{\frac23}}\epsilon^{nmkl}\psi_n\psi_m\psi_k\psi_l) 
\Bigr\}\ .
\end{equation}
The variables $e,x^n,\phi$ are even, the remaining ones are
odd. We regard all variables to be real except for $\bar\xi=\xi^*$. 
Reality of the kinetic term\footnote{
Alternatively, we could start with a conventional kinetic term 
$\bar\xi\dot\xi-\dot{\bar\xi}\xi$, which would lead to an equivalent 
quantum description in terms of creation/annihilation operators.} 
implies the boundary condition $\bar\xi\xi|^{\tau_2}_{\tau_1}=0$,
which restricts the trajectories entering the variational problem. 
The equations of motion in the $(\xi,\bar\xi)$ sector, 
\begin{equation}
\dot\xi+i\phi\xi\ =\ 0 \qquad{\rm and}\qquad 
\dot{\bar\xi}-i\phi\bar\xi\ =\ 0 \quad,
\end{equation}
can easily be integrated to yield
\begin{equation}\label{solvexi}
\xi\ =\ \xi_0\ e^{-i{\int_0}^\tau d\tau'\phi(\tau')} \qquad{\rm and}\qquad
\bar\xi\ =\ \bar{\xi}_0\ e^{i{\int_0}^\tau d\tau'\phi(\tau')} \quad,
\end{equation}
with a constant $\xi_0={({\bar\xi}_0)}^{*}$ 
conforming to the boundary condition.
{}From Eq.~(\ref{solvexi}) one observes
that only the constant modes $\xi_0$ and ${\bar\xi}_0$ 
enter the $\phi$ equation of motion,
\begin{equation}
i\bar\xi\xi-{\textstyle{\frac23}}\epsilon^{nmkl}\psi_n\psi_m\psi_k\psi_l
\ =\ 0 \quad.
\end{equation}
Since $(\xi,\bar\xi)$ do not appear elsewhere, this
is consistent with other equations of motion, in particular with
\begin{equation}
\epsilon^{nmkl}\dot\psi_n \psi_m\psi_k\psi_l\ =\ 0 \quad.
\end{equation}

Before going over to the Hamiltonian analysis we list the 
Lagrangian local world--line symmetries of the model:
\begin{eqnarray}
\delta x^n &=& \alpha\dot x^n\ 
	+\ \psi^n\epsilon\ 
	-\ \lambda\epsilon^{nmkl}\psi_m\psi_k\psi_l \quad,\nonumber\\[2pt]
\delta\psi^n &=& \alpha\dot{\psi^n}\  
	+\ {\textstyle{\frac{1}{2e}}}\Pi^n\epsilon\ 
	+\ {\textstyle{\frac{4}{3}}}\beta\epsilon^{nmkl}\psi_m\psi_k\psi_l\
	-\ {\textstyle{\frac{3}{2e}}}\lambda\epsilon^{nmkl}\Pi_m\psi_k\psi_l
	\quad,\nonumber\\[2pt]
\delta\xi &=& \alpha\dot\xi\ 
	-\ i\beta\xi \quad, \qquad\ \;
\delta\bar\xi\ =\ \alpha\dot{\bar\xi}\ 
	+\ i\beta\bar\xi \quad,\\[2pt]
\delta e &=& (\alpha e)^\cdot\ 
	+\ \chi\epsilon \quad, \qquad
\delta\phi\ =\ (\alpha\phi)^\cdot\ 
	+\ \dot\beta \quad,\nonumber\\[2pt]
\delta\chi &=& {(\alpha\chi)}^\cdot\ 
	+\ \dot\epsilon \quad, \qquad\ 
\delta\omega\ =\ {(\alpha\omega)}^\cdot\ 
	+\ {\textstyle{\frac{4}{3}}}\phi\epsilon\
	-\ {\textstyle{\frac{4}{3}}}\chi\beta\ 
	+\ \dot\lambda \quad,\nonumber
\end{eqnarray}
where we denote 
$\Pi^n\equiv\dot x^n+\omega\epsilon^{nmkl}\psi_m\psi_k\psi_l+\chi\psi^n$. 
In addition to world--line
reparametrization invariance and local supersymmetry of the massless 
spinning particle~(\ref{oldaction}), there appears a couple of new local 
symmetries with bosonic ($\beta$) and fermionic ($\lambda$) parameters
and their corresponding gauge fields $\phi$ and $\omega$, respectively. 
It is straightforward to check that the symmetry algebra closes 
only modulo the equations of motion.

Let us now analyze the system in the Hamiltonian formalism.
Evaluating the set of primary constraints\footnote
{We define momenta conjugate to anticommuting variables
to be right derivatives of the Lagrangian with respect to  
their velocities.}
\begin{eqnarray}\label{primary}
&&p_e=0 \quad, \qquad p_\phi=0 \quad, \qquad 
p_\omega=0 \quad,\qquad p_\chi=0 \quad, \nonumber\\
&&\quad p_{\psi n}-\psi_n=0 \quad, \qquad 
p_\xi-\bar\xi=0 \quad, \qquad p_{\bar\xi}=0 \quad,
\end{eqnarray}
one finds the canonical Hamiltonian to be
\begin{eqnarray}\label{hamiltonian}
H &=& p_e\lambda_e+p_{\phi}\lambda_\phi
+p_\omega\lambda_\omega
+(p_{\psi n}{-}\psi_n){\lambda_\psi}^n+p_\chi\lambda_\chi+
(p_\xi{-}\bar\xi)\lambda_\xi+p_{\bar\xi}\lambda_{\bar\xi}\nonumber\\[2pt]
&& +{\textstyle{\frac12}}e p^2-\omega\epsilon^{nmkl} p_n \psi_m\psi_k\psi_l-
\chi p^n\psi_n-\phi\,(i\bar\xi\xi-{\textstyle{\frac{2}{3}}}\epsilon^{nmkl}
\psi_n\psi_m\psi_k\psi_l) \quad,
\end{eqnarray}
where the $\lambda$'s are the Lagrange multipliers associated to
the primary constraints. Consistency conditions for the 
primary constraints imply the secondary ones,
\begin{eqnarray}\label{secondary}
p^n\psi_n &=& 0 \quad, \qquad 
i\bar\xi\xi - {\textstyle{\frac23}}\epsilon^{nmkl}
\psi_n\psi_m\psi_k\psi_l\ =\ 0 \quad,\nonumber\\[2pt]
p^2 &=& 0 \quad, \qquad\qquad\quad\ 
\epsilon^{nmkl} p_n \psi_m\psi_k\psi_l\ =\ 0 \quad, 
\end{eqnarray} 
and fix some of the Lagrange multipliers,
\begin{eqnarray}\label{fixmult}
{\lambda_\psi}^n &=& {\textstyle{\frac 12}} \chi p^n\
-\ {\textstyle{\frac 32}} \omega\epsilon^{nmkl} p_m \psi_k\psi_l\
+\ {\textstyle{\frac 43}} \phi \epsilon^{nmkl} \psi_m\psi_k\psi_l \quad,
\nonumber\\[2pt]
\lambda_\xi &=& -i\phi\xi \qquad{\rm and}\qquad 
\lambda_{\bar\xi}\ =\ i\phi\bar\xi \quad.
\end{eqnarray}
To show that no tertiary constraints arise at the next stage of the
Dirac procedure, one needs the well--known identity
\begin{equation}
\epsilon^{abcd}\epsilon_{efgd}\ =\
6\,{\delta^{[a}}_e{\delta^b}_f{\delta^{c]}}_g\ =\
6\,{\delta^a}_{[e}{\delta^b}_f{\delta^c}_{g]} \quad.
\end{equation}

The first line in Eqs.~(\ref{primary}) contains a set of first class 
constraints. Imposing the gauge
\begin{equation}\label{gauge}
e=1 \quad, \qquad \phi=0 \quad, \qquad \omega=0 \quad, \qquad \chi=0 
\end{equation}
allows us to omit four canonical pairs, the fixed
Lagrange multipliers being\footnote{
Eqs.~(\ref{gauge}) applied to Eqs.~(\ref{fixmult}) also imply
$\lambda_\xi=0=\lambda_{\bar\xi}$ 
which puts the Hamiltonian~(\ref{hamiltonian}) into its real form.
It was originally complex due to our choice for the $\xi$ kinetic term.}
\begin{equation}
\lambda_e=0 \quad, \qquad \lambda_\phi=0 \quad, \qquad 
\lambda_\omega=0 \quad, \qquad \lambda_\chi=0 \quad.
\end{equation}
Rewriting the second of Eqs.~(\ref{secondary}) in the equivalent form
\begin{equation}
ip_\xi\xi\ +\ i p_\xi p_{\bar\xi}\ -\ 
{\textstyle{\frac 23}}\epsilon^{nmkl}\psi_n\psi_m\psi_k\psi_l\ =\ 0 \quad,
\end{equation}
one decouples the (second class) constraints in the sector 
$(\bar\xi,p_{\bar\xi})$ from the others. 
In this fashion we may drop those variables, 
after having introduced the associated Dirac bracket.
The brackets for the remaining variables prove to be canonical. 
Finally, by making use of the shift 
\begin{equation}
\psi_n\ \longrightarrow\ \psi'_n\ =\
\psi_n+{\textstyle{\frac 12}} (p_{\psi n}-\psi_n) 
\qquad{\rm so\ that}\qquad
\{\psi'_n,\psi'_m\}\ =\ {\textstyle{\frac 12}}\eta_{nm} \quad,
\end{equation}
one can also isolate the second class constraint $p_{\psi n}{-}\psi_n=0$.
Regarding it further as a strong equation, one finally arrives at the 
following constraint set,
\begin{eqnarray}\label{constraints}
p^n\psi_n &=& 0 \quad, \qquad 
ip_\xi\xi-{\textstyle{\frac 23}}\epsilon^{nmkl} \psi_n\psi_m\psi_k\psi_l\ =\ 0
\quad, \nonumber\\[2pt]
p^2 &=& 0 \quad, \qquad\qquad\quad\ \
\epsilon^{nmkl} p_n \psi_m\psi_k\psi_l\ =\ 0 \quad, 
\end{eqnarray} 
which forms a closed algebra under the Dirac bracket.

Quantization is now straightforward. One promotes classical 
variables to quantum operators (here in position representation),
\begin{equation}
{\hat x}^n=x^n \quad, \qquad 
{\hat p}_n=-i\frac{\partial}{\partial x^n} \quad, \qquad
{\hat\psi}_n={\textstyle{\frac 12}}e^{\frac {3\pi i}{4}}\gamma_n \quad, \qquad
\hat\xi=\xi \quad, \qquad 
{\hat p}_\xi=i\frac{\partial}{\partial \xi} \quad,
\end{equation} 
which act on the corresponding quantum state represented by a wave function 
\begin{equation}\label{wave}
\Psi_A (x,\xi)\ =\ U_A (x)\ +\ \xi\, W_A (x) \quad.
\end{equation}
The latter is a Dirac spinor since it is acted upon by $R^{2,2}$ Dirac
matrices ${\gamma^n}_{AB}$.
In what follows, we assume the $qp$--ordering in the sector $(\xi,p_\xi)$.
In agreement with general principles we assign positive 
Grassmann parity to the $x$--dependent functions in the 
expansion~(\ref{wave}). Hence, the complete state has no definite parity.

Physical states are characterized by the fact
that the first class constraint operators annihilate them.
Eqs.~(\ref{constraints}) then yield 
\begin{equation}\label{fermion}
U(x)\ =\ 0 \quad,\qquad
(1+\gamma_5)W(x)\ =\ 0 \quad,\qquad
\gamma^n \partial_n W(x)\ =\ 0 \quad.
\end{equation}
At the quantum level, the second line in Eqs.~(\ref{constraints}) 
follows from the first one.

Thus, after quantization, the model~(\ref{newaction}) describes a single
on--shell chiral fermion. 

\bigskip

\noindent
{\bf 4. Adding rigid supersymmetry}\\
To obtain the remaining equations~(\ref{majorana}) from 
the quantization of our particle mechanics it suffices to 
extend the original configuration space $\{x^n\}$ 
to a superspace
$\{(x^n,\theta^\alpha,{\tilde\theta}^{\dot\alpha})\}$, 
by adjoining a pair $(\theta^\alpha,{\tilde\theta}^{\dot\alpha})$ 
of Kleinian Majorana--Weyl spinors, 
and to provide a couple of first class constraints 
\begin{equation}\label{couple}
p_{\bar\theta\dot\alpha}-\theta^\alpha p_{\alpha\dot\alpha}\ =\ 0 
\qquad{\rm and}\qquad
p_{\theta\alpha}\,p^{\alpha\dot\alpha}\ =\ 0 
\end{equation}
when passing to the Hamiltonian formalism. 
This is achieved by extending the action~(\ref{newaction}) 
just like in the construction of Siegel's 
superparticle~\cite{siegel1} (see also related works~\cite{gala1,gala2}),
\begin{eqnarray}
S&=&\int^{\tau_2}_{\tau_1}\!\! d\tau \Bigl\{ {\textstyle{\frac 1{2e}}} 
(\dot x^n-i\dot\theta\sigma^n\tilde\theta+i\theta\sigma^n\dot{\tilde\theta}
+\omega\epsilon^{nmkl}\psi_m\psi_k\psi_l
+\chi\psi^n+i\rho\sigma^n\tilde\mu)^2 
\nonumber\\[2pt] &&\qquad
+\ \psi^n {\dot\psi}_n +\bar\xi\dot\xi-\rho\,{\dot\theta}
+\phi\, (i\bar\xi\xi 
-{\textstyle{\frac23}}\epsilon^{nmkl}\psi_n\psi_m\psi_k\psi_l)
\Bigr\} \quad.
\end{eqnarray}
Here, $\rho^\alpha$ and ${\tilde\mu}^{\dot\alpha}$ are auxiliary 
(real) odd variables.   We switch back to spinor notation.
The model is invariant under 
rigid space--time supersymmetry transformations
\begin{equation}
\delta x^{\alpha\dot\alpha}\ =\ 
2(\theta^\alpha\tilde\epsilon^{\dot\alpha}
-\epsilon^\alpha\tilde\theta^{\dot\alpha}) 
\quad, \qquad \delta\theta^\alpha\ =\ \epsilon^\alpha \quad, \qquad 
\delta\tilde\theta^{\dot\alpha}\ =\ \tilde\epsilon^{\dot\alpha} \quad. 
\end{equation}
The set of local symmetries is extended by
a pair of new symmetries, including kappa symmetry,
\begin{eqnarray}
\delta\theta^\alpha &=& 
-e^{-1} \Pi^{\alpha\dot\alpha}\tilde\kappa_{\dot\alpha} \quad,\qquad
\delta\tilde\theta^{\dot\alpha}\ =\ 0 \quad,\qquad
\delta x^{\alpha\dot\alpha}\ =\ 
2(\delta\theta^\alpha\tilde\theta^{\dot\alpha} 
-\rho^\alpha\tilde\kappa^{\dot\alpha}) \quad,
\nonumber\\[2pt]
\delta\rho^\alpha &=& 0 \quad,\qquad\qquad\qquad\ \;
\delta\tilde\mu^{\dot\alpha}\ =\ 
\dot{{\tilde\kappa}^{\dot\alpha}}\quad,\qquad\ \;
\delta e\ =\ -4\,\tilde\kappa_{\dot\alpha} \dot{{\tilde\theta}^{\dot\alpha}}
\end{eqnarray}
and
\begin{eqnarray}
\delta\theta^\alpha &=& 0 \quad,\qquad\qquad\qquad\quad
\delta\tilde\theta^{\dot\alpha}\ =\ \tilde\nu^{\dot\alpha} \quad,\quad\ \,
\delta x^{\alpha\dot\alpha}\ =\ -2\,\theta^\alpha\tilde\nu^{\dot\alpha} \quad,
\qquad\quad \nonumber\\[2pt]
\delta\rho^\alpha &=& 
2\, e^{-1} \Pi^{\alpha\dot\alpha}\tilde\nu_{\dot\alpha} \quad,\qquad\ \,
\delta\tilde\mu^{\dot\alpha}\ =\ 0 \quad,\qquad\quad\;
\delta e\ =\ 4\,\tilde\nu_{\dot\alpha} \tilde\mu^{\dot\alpha} \quad,
\end{eqnarray}
where 
\begin{equation}
\Pi^{\alpha\dot\alpha}\ =\ \dot x^{\alpha\dot\alpha}
-2\,\dot\theta^\alpha\tilde\theta^{\dot\alpha}
+2\,\theta^\alpha\dot{{\tilde\theta}^{\dot\alpha}}
+\omega\psi^{\alpha\dot\beta}\psi_{\beta\dot\beta}\psi^{\beta\dot\alpha}
+\chi\psi^{\alpha\dot\alpha}
+2\,\rho^\alpha\tilde\mu^{\dot\alpha} \quad.
\end{equation}
As is seen from the transformations 
above, the $\tilde\nu$ symmetry allows us to remove the
$\tilde\theta$ variable, which is compatible with the first of the
first class constraints~(\ref{couple}) 
arising in the Hamiltonian formalism.

The Hamiltonian analysis for the extended model proceeds along 
the lines of the previous section. 
In addition to the seven earlier primary constraints~(\ref{primary}),
one finds four new ones,
\begin{equation}\label{sprimary}
p_{\rho\alpha}\ =\ 0 \quad,\qquad
p_{\tilde\theta\dot\alpha}-\theta^\alpha p_{\alpha\dot\alpha}\ 
=\ 0 \quad,\qquad
p_{\tilde\mu\dot\alpha}\ =\ 0 \quad,\qquad
p_{\theta\alpha}-\tilde\theta^{\dot\alpha} p_{\alpha\dot\alpha}-\rho_\alpha\
=\ 0 \quad,
\end{equation}
while the Hamiltonian~(\ref{hamiltonian}) acquires the contribution
\begin{equation}
H_{add}\ =\ 
p_{\rho\alpha}{\lambda_\rho}^\alpha+
p_{\tilde\mu\dot\alpha}{\lambda_{\tilde\mu}}^{\dot\alpha}+
(p_{\theta\alpha}-\tilde\theta^{\dot\alpha} p_{\alpha\dot\alpha}-\rho_\alpha)
{\lambda_\theta}^\alpha+
(p_{\tilde\theta\dot\alpha}-\theta^\alpha p_{\alpha\dot\alpha})
{\lambda_{\tilde\theta}}^{\dot\alpha}-
\rho^\alpha \tilde\mu^{\dot\alpha} p_{\alpha\dot\alpha} \quad.
\end{equation}
Time translation invariance of the new constraints~(\ref{sprimary}) 
yields a new secondary one as well,
\begin{equation}\label{ssecondary}
\rho^\alpha p_{\alpha\dot\alpha}\ =\ 0 \quad,
\end{equation}
and specifies some of the Lagrange multipliers, i.e.  
\begin{equation}
\lambda_{\theta\alpha}\ =\  
\tilde\mu^{\dot\alpha} p_{\alpha\dot\alpha} \quad,\qquad
\lambda_{\rho\alpha}\ =\ 
-2\,{\lambda_{\tilde\theta}}^{\dot\alpha} p_{\alpha\dot\alpha} \quad.
\end{equation}
No tertiary constraints appear at the next stage of the Dirac procedure. 
In the sector of the additional variables 
$(\theta,\tilde\theta,\rho,\tilde\mu)$
the full set of constraints may then be written as
\begin{equation}\label{second}
p_{\theta\alpha}-\tilde\theta^{\dot\alpha}p_{\alpha\dot\alpha}-\rho_\alpha\
=\ 0 \quad,\qquad p_{\rho\alpha}\ =\ 0 \quad,
\end{equation}
\begin{equation}\label{first}
(p_{\theta\alpha}-\tilde\theta^{\dot\alpha}p_{\alpha\dot\alpha})
p^{\alpha\dot\beta}\ =\ 0 \quad,\qquad 
p_{\tilde\theta\dot\alpha}-\theta^\alpha p_{\alpha\dot\alpha}
+2\,{p_\rho}^\alpha p_{\alpha\dot\alpha}\ =\ 0 \quad,\qquad
p_{\tilde\mu\dot\alpha}\ =\ 0 \quad,
\end{equation}
The constraints~(\ref{first}) are first class. Imposing the
gauge (implying $\lambda_{\tilde\mu}=0$)
\begin{equation} 
{\tilde\mu}^{\dot\alpha}\ =\ 0 \quad,
\end{equation} 
one can omit the variables $(\tilde\mu,p_{\tilde\mu})$. 
The pair~(\ref{second}) 
is second class and can be taken as strong equations 
after introducing the associated Dirac bracket.
The Dirac brackets for the remaining variables turn out to
coincide with the Poisson ones. Upon quantization,
the remaining constraints will append two additional 
restrictions to equations~(\ref{fermion}), namely 
\begin{equation}
\tilde D_{\dot\alpha} W_A (x,\theta,\tilde\theta)\ =\ 0 
\qquad{\rm and}\qquad
\partial^{\alpha\dot\alpha} D_\alpha W_A (x,\theta,\tilde\theta)\ =\ 0 \quad,
\end{equation}
with $W_A$ from~(\ref{fermion}) acquiring now 
$(\theta,\tilde\theta)$ dependence. 
Beautifully enough, these are precisely the equations describing 
abelian self--dual super Yang--Mills in their complexified form.\footnote{
In $R^{1,3}$, time reversal is often used to impose a reality condition
on the wave function~\cite{howe}, which is, of course, only admissible
for energy eigenfunctions. For $R^{2,2}$ however, this is not an option
because there is no notion of orthochronicity.}

\vfill\eject
\noindent
{\bf 5. Concluding remarks}\\
To summarize, in this letter we have constructed a particle 
model which describes (abelian) SDSYM upon quantization.
The theory is essentially doubly supersymmetric, i.e. combines
the features of the Green--Schwarz and Neveu--Schwarz--Ramond 
formalisms. In order to attempt a superstring generalization, it seems 
natural to start with a string analog of the action~(\ref{newaction}), 
followed by the implementation of kappa symmetry. We envisage 
that the fermionic oscillator degree of freedom, being inessential in 
the particle case, may exhibit a nontrivial influence on the string spectrum.
Another (heterotic) possibility is provided by combining the standard 
NSR set of constraints in the left-moving sector with the constraints 
of Ref.~\cite{pope1} in the right-moving sector. Such an approach is likely
to be related to that of Ref.~\cite{deboer}.

As has been discussed in Ref.~\cite{vanholten2}, the global symmetry structure 
underlying a wide class of doubly supersymmetric particles is
that of the superconformal group. It would be interesting to  
perform a similar analysis for the formulation at hand. 

\bigskip
\noindent
{\bf Acknowledgments}\\
A.G. thanks S.V. Ketov and S.M. Kuzenko for useful discussions 
and the DAAD for financial support.

\end{document}